\documentstyle[preprint,aps]{revtex}
 
\begin{document}
\draft
\title{Quantum cavitation in liquid helium}
\author{Montserrat Guilleumas$^1$, Manuel Barranco$^2$,
Dora M. Jezek$^3$, Roland J. Lombard$^4$, and Mart\'{\i} Pi$^2$}
\address{$^1$Dipartimento di Fisica, Universit\`a di Trento. 38050
Povo, Italy}
\address{$^2$Departament d'Estructura i Constituents de la Mat\`eria,
Facultat de F\'{\i}sica, \\
Universitat de Barcelona, E-08028 Barcelona, Spain}
\address{$^3$Departamento de F\'{\i}sica, Facultad de Ciencias Exactas
y Naturales, \\
Universidad de Buenos Aires, RA-1428 Buenos Aires, Argentine}
\address{$^4$Institut de Physique Nucl\'eaire, Division de Physique
Th\'eorique. 91406 Orsay, France}

\maketitle
 
\begin{abstract}
 
Using a functional-integral approach, we have determined the
temperature below which cavitation in liquid helium is driven by
thermally assisted quantum tunneling. For both helium isotopes, we
have obtained the crossover temperature in the whole range of allowed
negative pressures. Our results are compatible with recent
experimental results on $^4$He.

\end{abstract}
 
 
%\section{Introduction}
 
The possibility of having observed quantum cavitation in superfluid
$^4$He has been first put
forward by Balibar and coworkers \cite{Bal1}. These authors have used
a
hemispherical transducer that focusses a sound wave in a small region
of a cell where cavitation is induced in liquid $^4$He at low
temperature. The analysis of their experimental data is
complicated by the fact that neither the pressure (P) nor the
temperature (T) at the focus can
be directly measured. This makes the determination of the
thermal-to-quantum
cavitation crossover temperature T$^*$ to depend on the theoretical
equation of state (EOS) near the spinodal point. Using the results of
Ref. \cite{Maris1}, they conclude that T$^* \sim$ 200 mK, in agreement
with the prediction of \cite{Maris1}. However,  using for instance
the EOS of Ref. \cite{Guirao}, which reproduces the spinodal point
microscopically calculated by Boronat et al \cite{Boronat1,Boronat2},
the "experimental" result becomes 120 mK.
%, in apparently better
%agreement with the prediction of Ref. \cite{Guilleumas1}.
 
The first detailed description of the cavitation process in liquid
helium was provided by Lifshitz and Kagan \cite{Lif1}, who used the
classical capillarity model near the saturation line, and a density
functional-like description near the spinodal line. More recently,
the method has been further elaborated by Xiong and Maris \cite{Xiong}.
These authors conclude that there is no clear way to interpolate
between these two regimes, which makes quite uncertain the range of
pressures in which each of them is valid.
 
  In this work, we determine T$^*$ for $^3$He and $^4$He using a
functional-integral approach (FIA) in conjunction with a density
functional description of liquid helium. The method overcomes the
conceptual limitations of previous works based on the application of
 zero-temperature
multidimensional WKB methods \cite{Maris1}, and the technical ones
inherent to the use of parametrized bubble density profiles
\cite{Guilleumas1}, thus putting on firmer grounds the
theoretical results. Moreover, it gives T$^*$ in the whole pressure
range.

   Thermally assisted quantum tunneling is nowadays well
understood (see for example Ref. \cite{Chud} and Refs. therein). Let
us simply recall that at high temperatures, the cavitation rate, i.e.,
the number of bubbles formed per unit time and volume, is given by
\begin{equation}
J_T = J_{0T}\, e^{-\Delta\Omega_{max}/T}\, ,
\label{eq1}
\end{equation}
where $\Delta\Omega_{max}$ is the barrier height for thermal
activation and $J_{0T}$ is a prefactor which depends on the dynamics
of the cavitation process. At low T, it becomes
\begin{equation}
J_Q = J_{0Q}\, e^{-S_{min}}\, ,
\label{eq2}
\end{equation}
where $S_{min}$ is the minimum of the imaginary-time action
\begin{equation}
S(T) =  \oint {\rm d}\tau \int {\rm d}{\vec r}\,\, {\cal L}\, ,
\label{eq3}
\end{equation}
${\cal L}$ being the imaginary-time classical Lagrangian density of
the system and the time-integration is extended over a period in the
potential well obtained by inverting the potential barrier. These
equations hold provided the rate can be calculated in the
semiclassical limit, i.e., $S_{min} >> 1$, which is the present case.
For a given value of T, one has to obtain periodic solutions to the
variational problem embodied in Eq. (\ref{eq3}). Among these many
periodic solutions, called thermons in Ref. \cite{Chud}, those
relevant for the problem of finding T$^*$ are the ones corresponding
to small oscillations around the minimum of the potential, which has
an energy equal to $-\Delta\Omega_{max}$. If $\omega_p$ is the angular
frequency of this oscillation, T$^*=\hbar\omega_p/2\pi$. It is worth
realizing that contrarily to WKB, this procedure
permits to go continously from one regime to the other: at T$^*$, Eqs.
(\ref{eq1}) and (\ref{eq2}) coincide, whereas the WKB approach forces
to equal a zero-temperature barrier penetrability to a
finite-temperature Arrhenius factor
\cite{Maris1,Guilleumas1}. Whether this is justified or not, can only
be ascertained a posteriori  comparing the WKB with FIA
results.
 
   To obtain the Lagrangian density ${\cal L}$ we have resorted to a
zero-temperature density functional description of the system
\cite{Guirao,Guilleumas2}. This is justified in view of the low-T
that are expected to come into play ($\le$ 200 mK). The critical
cavity
density profile $\rho_0(r)$ is obtained  solving the Euler-Lagrange
equation \cite{Xiong,Jezek}
\begin{equation}
\frac{\delta\omega} {\delta\rho} = 0\, ,
\label{eq4}
\end{equation}
where $\omega(\rho)$ is the grand potential density and $\rho$ is the
particle density. $\Delta\Omega_{max}$ is given by
\begin{equation}
\Delta\Omega_{max} =\int {\rm d}{\vec
r}\left[\omega(\rho_0)-\omega(\rho_m)\right]\, ,
\label{eq5}
\end{equation}
where $\rho_m$ is the density of the metastable homogeneous liquid. It
is now simple to describe the dynamics of the cavitation process in
the
inverted barrier well, whose equilibrium configuration corresponds to
$\rho_0(r)$ and has an energy $-\Delta\Omega_{max}$. We suppose that
the collective velocity of the fluid associated with the bubble growth
is irrotational. This is not a severe restriction since one expects
only radial displacements (spherically symmetric bubbles). Introducing
the velocity potential field s$({\vec r},t)$, we have
\begin{equation}
%{\cal L} =m\rho \dot{s} - {\cal H}(\rho, s)\, ,
{\cal L} =m\dot{\rho}s - {\cal H}(\rho, s)\, ,
\label{eq6}
\end{equation}
where ${\cal H}(\rho,s)$ is the imaginary-time hamiltonian density.
Defining ${\vec u}({\vec r},t)\equiv \nabla s({\vec r},t)$,
\begin{equation}
{\cal H} =\frac{1}{2}m\rho{\vec u}^2
-\left[\omega(\rho)-\omega(\rho_m)\right]\,\, .
\label{eq7}
\end{equation}
Hamilton's equations yield
\begin{equation}
m\dot{\rho} = \frac{\delta {\cal H}} {\delta s} = -m \nabla(\rho
{\vec u})
\label{eq8}
\end{equation}
\begin{equation}
m\dot{s} = -\frac{\delta{\cal H}} {\delta\rho}\, .
\label{eq9}
\end{equation}
Eq. (\ref{eq8}) is the continuity equation. Taking the gradient of Eq.
(\ref{eq9}) we get
\begin{equation}
m\frac{{\rm d}{\vec u}}{{\rm dt}} = -\nabla
\left\{
\frac{1}{2}m{\vec u}\,^2-\frac{\delta\omega}{\delta\rho}
\right\}\, .
\label{eq10}
\end{equation}

Thermons $\rho({\vec r},t)$ are periodic solutions of Eqs. (\ref{eq8})
and (\ref{eq10}). From Eq. (\ref{eq3}) and using Eqs. (\ref{eq6}) and
(\ref{eq8}) we can write
\begin{equation}
S_{min}(T) = \oint {\rm d}\tau \int {\rm d}{\vec r}
\left\{
\frac{1}{2}
%m{\vec u}\,^2+\frac{\delta\omega}{\delta\rho}
m\rho{\vec u}^2+\omega(\rho)-\omega(\rho_m)
\right\}\, .
\label{eq11}
\end{equation}
Within this model, to {\it exactly} obtain T$^*$ only a linearized
version
of Eqs. (\ref{eq8}) and (\ref{eq10}) around $\rho_0(r)$ is needed.
Defining the T$^*$-thermon as
\begin{equation}
\rho(r,t) \equiv \rho_0(r) + \rho_1(r)\, e^{i\omega_p t}\, ,
\label{eq12}
\end{equation}
where $\rho_1(r)$ is much smaller than $\rho_0(r)$, and keeping only
first order terms in ${\vec u}(r,t)$ and $\rho_1(r)$, we get:
\begin{equation}
m\omega_p^2 \rho_1(r)= \nabla\left[\rho_0(r) \nabla\left(
\frac{\delta^2\omega}{\delta\rho^2}\bullet\rho_1(r)\right)\right]\, .
\label{eq13}
\end{equation}
Here, $\frac{\delta^2\omega}{\delta\rho^2}\bullet\rho_1(r)$ means that
$\delta\omega / \delta\rho$ has to be linearized, keeping only
terms in $\rho_1(r)$ and its derivatives.
 
   Eq. (\ref{eq13}) is a fourth-order linear differential, eigenvalue
equation. A careful analysis shows that its
  physical solutions have to fulfill
$\rho_1'(0)=\rho_1'''(0)=0$, and fall
exponentially
to zero at large distances. The linearized continuity equation
$\rho_1(r)\propto -\nabla(\rho_0{\vec u})$ imposes the integral of
$\rho_1(r)$ to yield zero when taken over the whole space.
 
  We have solved Eq. (\ref{eq13})  using seven point Lagrange
formulae to discretize the r-derivatives together with a standard
diagonalization
subroutine. The sensibility of the solution to the precise value of
the
r-step has been carefully checked, and in most cases a value $\Delta
r$ = 0.25 \AA\, has been used.
 
 For all pressures, only one positive
$m\omega_p^2$ eigenvalue has been found. Fig. 1 (a) and (b) shows
T$^*$
(mK) as a function of P(bar) for $^4$He and $^3$He, respectively. In
the case of $^4$He, the maximum T$^*$ is 238 mK at -8.58 bar, and for
$^3$He it is 146 mK at -2.91 bar. It is worth noting that T$^*$ is
strongly dependent on P in the spinodal region, falling to zero at the
spinodal point (see also Ref. \cite{Xiong}).
 
 We  display in
Fig. 2 the $\rho_1(r)$-component of the thermon (\ref{eq12}) in the
case of $^4$He (a similar figure could be drawn for $^3$He). For large
bubbles, $\rho_1(r)$ is localized at the surface: the thermon is a
well defined surface excitation.
It justifies the use of the capillarity approximation
 near saturation, or more elaborated approaches,
 like that of Ref.
 \cite{Guilleumas1}, that consists in a simplified one-dimensional
model in which the oscillations are just described by rigid
 displacements of the critical bubble surface.
 
 When the density inside the bubble becomes sizeable, a mixed
surface-volume thermon develops, which eventually becomes a pure
volume
mode in the spinodal region. This mode can no longer be described as a
rigid density displacement, and the above mentioned models fail:
 the exact T$^*$ is higher than the prediction of the rigid surface
displacement model because volume modes involve higher frequencies.
 
  To determine which of the T$^*$(P) shown in Fig. 1 corresponds to
the actual experimental conditions, we have calculated the homogeneous
cavitation pressure P$_h$ \cite{Xiong,Jezek}. It is the one
the system can sustain before bubbles nucleate at an appreciable
rate. We have solved the equation
\begin{equation}
1 = (Vt)_e\, J
\label{eq14}
\end{equation}
taking J$=$J$_T$ and
\begin{equation}
J_{0T} = \frac{k_B T}{h V_0}.
\label{eq14b}
\end{equation}
\noindent $V_0 = 4 \pi R^3_c/3$ represents the volume of the critical
bubble, for which we have  taken R$_c = $10 \AA. For T $<$
T$^*$,  J$_{T}$ has to be replaced by J$_{Q}$. Lacking of a
better choice, we have taken  J$_{0Q} =$ J$_{0T}$(T$=$T$^*$), and
  for the experimental factor
(Vt)$_e$ (experimental volume$\times$time), two values at the limits of
the experimental range
\cite{Bal1,Maris1}, namely 10$^{14}$ and 10$^4$ \AA$^3$ s. For
$^4$He it yields P$_h$=-8.57 bar and -8.99 bar, respectively. The
corresponding values for $^3$He are -2.97 and -3.06 bar. This means
that for both isotopes P$_h$ is close to the spinodal pressure. Table
1 displays the associated T$^*$-values.
 
    The crossover temperatures are similar to those
given in Ref. \cite{Maris1}, although different functionals
have been used in both calculations.
As a matter of fact,  this is irrelevant, since
both functionals reproduce equally well the  experimental
quantities  pertinent to the description of the cavitation process.
 
An explanation for the agreement between these calculations can be
found in Ref. \cite{Guilleumas1}. In that work, using a simplified
one-dimensional model in which the oscillations were modelled by rigid
displacements of the bubble surface,
the cavitation process was described within FIA from
T=0 to the thermal regime. It was shown that thermally assisted
quantum cavitation only adds small corrections to the T=0 "instanton"
solution (formally equivalent to WKB if S$_{min}>>1$) in the
quantum-to-thermal transition region.
 
 Let us recall that the formalism used in Ref. \cite{Maris1} to
describe quantum cavitation is a multidimensional WKB one, appropiated
for a T=0, pure quantum state with a well defined energy value. This
approximation is well known to fail for energies close to the
top of the barrier. On the contrary, the FIA here adopted deals with
thermally mixed quantum states, making it possible to smoothly connect
quantum and thermal regimes \cite{Chud}.
 Besides, it is
technically complicated to obtain the  E=0 instanton
solution to Eqs. (\ref{eq8}) and (\ref{eq10}) without using some
numerical approximations
\cite{Maris1} that might be unworkable in more complex physical
situations, like that of a $^3$He-$^4$He liquid mixture.
We also want to stress again that, to determine the quantity of
experimental significance, namely T$^*$,
 only the thermon solution of the much simpler
eigenvalue Eq. (\ref{eq13}) is required.

   To conclude, within density functional theory,
   we have performed a thorough description of the
quantum-to-thermal transition in the process of cavitation in liquid
helium based on the
functional-integral approach. Our quantitative results (see also Ref.
\cite{Maris1}) indicate that the crossover temperature is below 240 mK
for $^4$He, and below 150 mK for $^3$He. The experiments on
$^4$He yield results
which, depending on which equation of state is used, are in the
120-200 mK range. Given the present uncertainties in theoretical and
experimental results as well, we consider the agreement as
satisfactory.
 
   We would like to thank Sebastien Balibar,
Eugene Chudnovsky and Jacques Treiner
for useful discussions. This work has been supported by DGICYT
(Spain) Grant No. PB92-0761, by the Generalitat de Catalunya Grant No.
GRQ94-1022, by the CONICET (Argentine) Grant No. PID 97/93 and by the
IN2P3-CICYT agreement.

\begin{figure}
\caption{
(a) T$^*$(mK) as a function of P(bar) for $^4$He. (b) Same as (a)
for $^3$He.
}
\label{fig1}
\end{figure}
\begin{figure}
\caption{
Referring to $^4$He, we show: (a) the particle density profile
$\rho_0(r)$ (solid line) and the density $\rho_1(r)$ (dashed line) for
P=-4.59 bar.
(b) Same as (a) for P=-8.35 bar.
(c) Same as (a) for P=-9.16 bar.
$\rho_1(r)$ is drawn in arbitrary units, $\rho_0(r)$ in
\AA$^{-3}$ and $r$ in \AA.\
}
\label{fig2}
\end{figure}
\begin{table}
\caption{
Crossover temperatures for two different values of the
experimental volume times time.
}
\begin{tabular}{ccc}
(Vt)$_e$\,(\AA$^3$ s)&\multicolumn{2}{c}{T$^*$(mK)}\\
\cline{2-3}
&$^3$He&$^4$He\\
\tableline
10$^{14}$&143&238\\
10$^4$&106&198\\
\end{tabular}
\label{table1}
\end{table}
 

\begin{references}
\bibitem{Boronat2} The result for the spinodal pressure of
\cite{Guirao} is P=-9.08 atm at $\rho$=0.0159 \AA$^{-3}$ as
compared with that of \cite{Boronat1}, P=-9.30$\pm$0.15 atm at
$\rho$=0.0158$\pm$0.0001 \AA$^{-3}$.
\end{references}
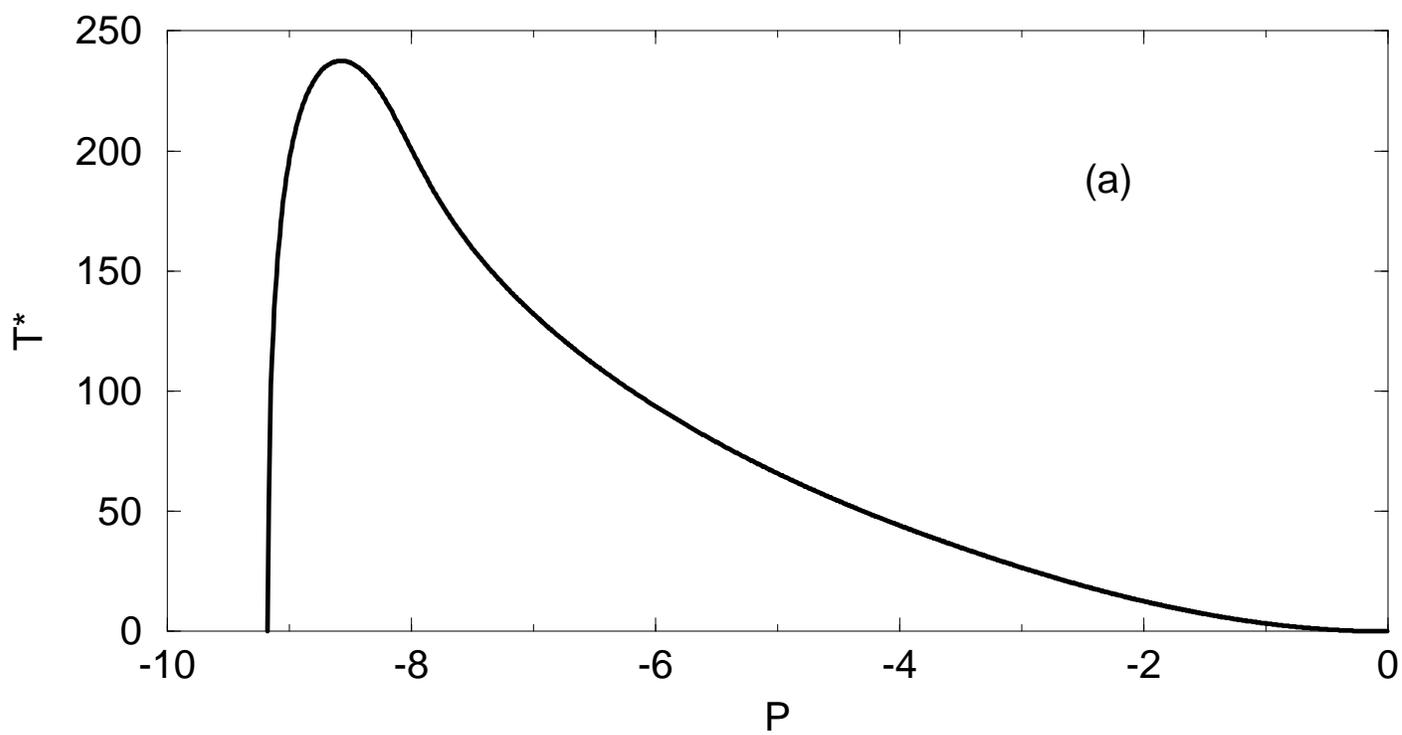
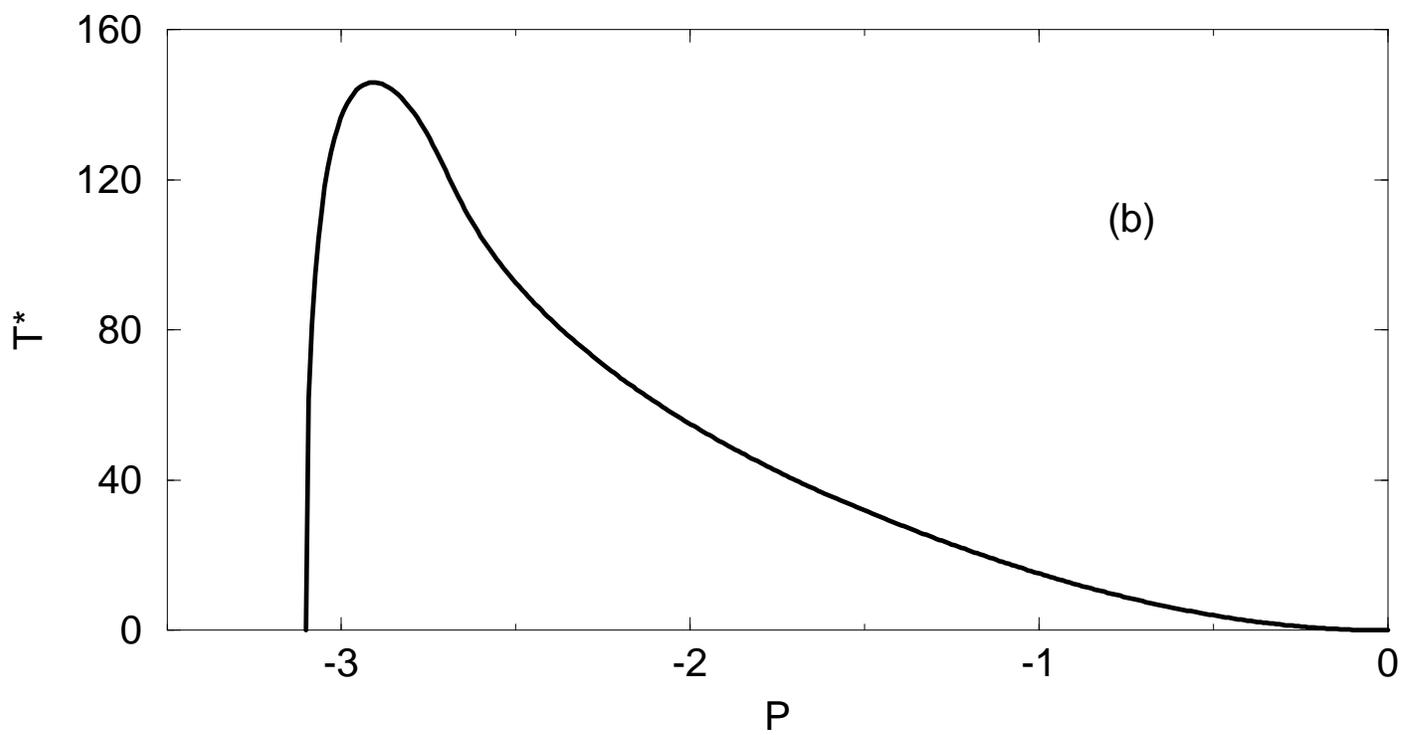
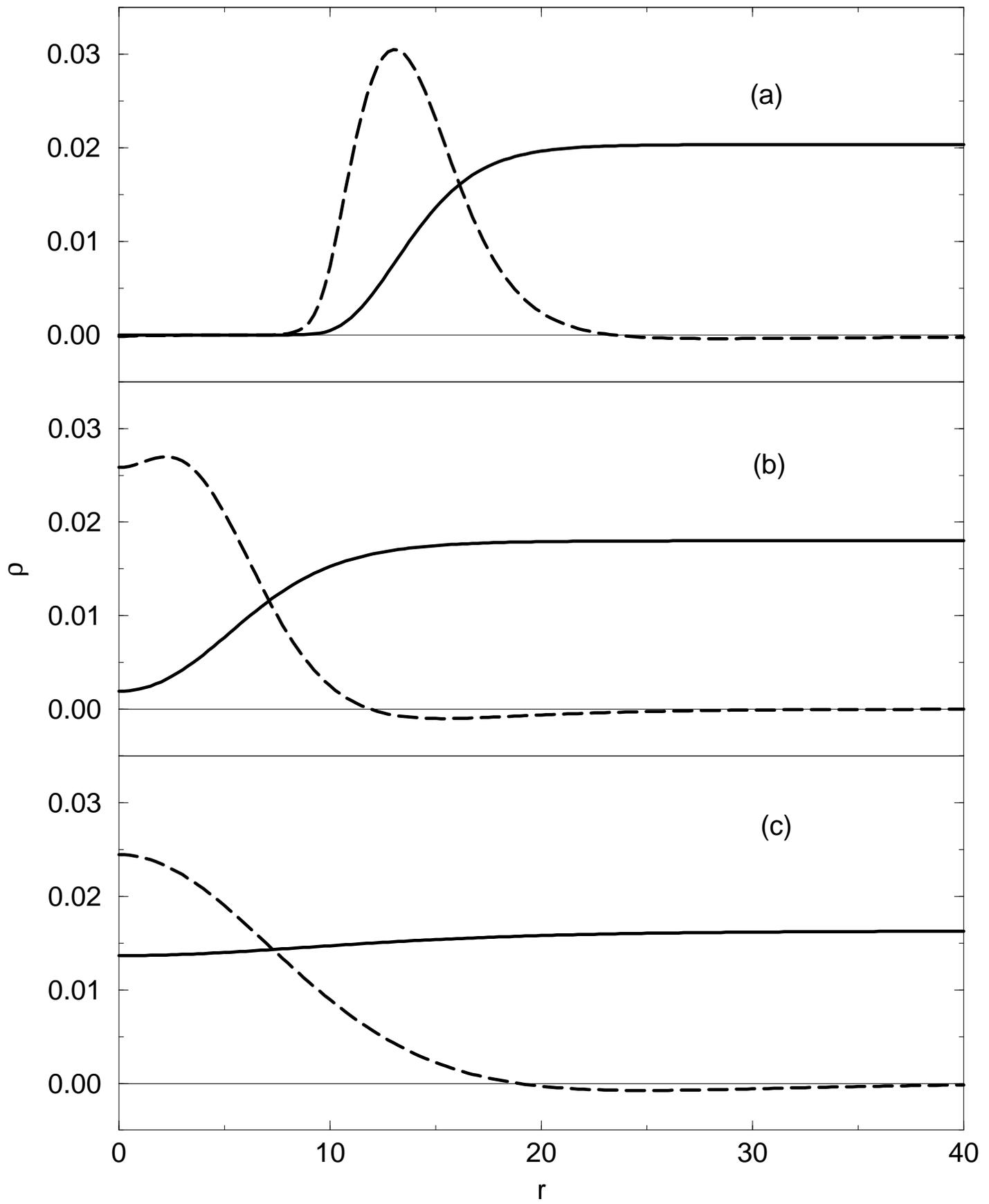
\end{document}